\documentclass[conference, 9pt]{IEEEtran}
\IEEEoverridecommandlockouts
\usepackage[backend = biber, style=ieee, maxnames = 6, minnames = 1]{biblatex}
\AtBeginBibliography{\small}
\usepackage{amsmath,amssymb,amsfonts}
\usepackage{amsthm}
\usepackage{algorithmic}
\usepackage{graphicx}
\usepackage{bbm}
\usepackage{subfig}
\usepackage{textcomp}
\usepackage{booktabs}
\usepackage{xcolor}
\usepackage{qcircuit}
\def\BibTeX{{\rm B\kern-.05em{\sc i\kern-.025em b}\kern-.08em
    T\kern-.1667em\lower.7ex\hbox{E}\kern-.125emX}}
\begin{document}

\title{Fault Models in Superconducting Quantum Circuits\\
}

\author{\IEEEauthorblockN{Qifan Huang}
\IEEEauthorblockA{\textit{State Key Laboratory of Computer Science} \\
\textit{Institute of Software, Chinese Academy of Science}\\
\textit{University of Chinese Academy of Science} \\
Beijing, China\\
huangqf@ios.ac.cn}
\and
\IEEEauthorblockN{Boxi Li}
\IEEEauthorblockA{ \textit{Institute of Quantum Control} \\
\textit{Forschungszentrum Jülich} \\ 
Jülich D-52425, German\\
b.li@fz-juelich.de}
\and
\IEEEauthorblockN{Minbo Gao}
\IEEEauthorblockA{\textit{State Key Laboratory of Computer Science}\\
\textit{Institute of Software, Chinese Academy of Science}\\
\textit{University of Chinese Academy of Science}\\
Beijing, China\\
gaomb@ios.ac.cn}
\and
\IEEEauthorblockN{Mingsheng Ying}
\IEEEauthorblockA{\textit{State Key Laboratory of Computer Science} \\
\textit{Institute of Software, Chinese Academy of Science}\\
\textit{Department of Computer Science, Tsinghua University} \\
Beijing, China\\
yingms@ios.ac.cn}
}

\maketitle

\begin{abstract}

Fault models are indispensable for many EDA tasks, so as for design and implementation of quantum hardware. In this article, we propose a fault model for  superconducting quantum systems. Our fault model reflects the real fault behavior in control signals and structure of quantum systems.
Based on it, we conduct fault simulation on controlled-Z gate and quantum circuits by QuTiP. We provide fidelity benchmarks for incoherent faults and test patterns of minimal test repetitions for coherent faults. Results show that with 34 test repetitions a 10\% control noise can be detected, which help to save test time and memory.

%Fault models are indispensable for many EDA tasks (e.g. fault simulation and ATPG), so as for the design and implementation of quantum hardware. In this article, we propose fault models for one of the most promising quantum hardware technology -- superconducting quantum systems. Our fault model reflects the Based on it, we conduct fault simulation on two-qubit controlled-Z gate and quantum circuits with the aid of QuTiP. 
%--  a popular software for simulating the dynamics of open quantum systems. 
%Our results show that a maximum noise ratio $20\%$ corresponds to a fidelity loss up to $6\%$. Moreover, simulation on a 4-qubit quantum full adder with a $10\%$ noise inserted in the control pulse in every CZ gate the minimum test repetition is 34 while the noise inserted in only one CZ gate leads to fault either being undetectable or being detected in between $100-200$ times. 

\end{abstract}


\section{Introduction}
Quantum computing is considered as one of the most promising next-generation computing technology. Over the past few decades, significant progress has been made in implementing quantum computing by real physical systems such as quantum dots, ion traps and superconducting systems. Among them, superconducting quantum systems are widely acknowledged as one of the most promising technologies. Qubits made of superconducting devices harness the advantages of scalability and stability. 
%, and has been used to demonstrate some algorithms in the 'noisy intermediate scale quantum'(NISQ) era. Lots of achievements have been made in making a superconducting quantum computer, including Google's 53 qubits Sycamore \cite{2019Quantum} and USTC's 62 qubits $8\times8$ quantum chip  \cite{gong2021quantum}. 
%Theory has proved that single qubit gates and two qubit controlled-not (CNOT) gates form a universal gate set. Moreover,in the past ten years, 
In particular, two qubit gates, e.g. controlled-Z (CZ) gates, have been realized in superconducting quantum systems with the optimal fidelity known as $99.76\% \pm 0.07\%$ \cite{PhysRevX.11.021058}.
%, therefore ensuring the development of quantum design automation (QDA). However, quantum algorithms based on 

Real superconducting quantum devices suffer from several serious noises, including leakage, decoherence, and control pulse noises. Therefore, proper fault models of quantum gates are indispensable in various tasks of design automation for superconducting quantum circuits, e.g. 
%of quantum gates must be considered when performing verification, 
testing, simulation and verification. 
%and logic synthesis in QDA process. 
Indeed, several faults of superconducting quantum circuits have been considered in physical experiments, and some measures to suppress these faults have been proposed. However, there still lacks an appropriate fault model for superconducting quantum circuits that is suited to various design automation issues (e.g. fault simulation and automatic test pattern generation (ATPG)).
%, and simulation benchmarks based on virtual platforms. Therefore, developing an overall fault model is an necessity. 

In this work, we introduce such a fault model, which includes both coherent faults and incoherent faults. Roughly speaking, coherent faults originate from the imperfect control signals while incoherent faults are due to the interaction with environments. To test the effectiveness of our model, we conduct fault simulation on single controlled-Z gates and certain quantum circuits, using QuTiP \textemdash \ a popular
software for simulating quantum systems \textemdash \ as the main simulation tool. The simulation results show that with the undesired rotation inserted in the CZ gate, the fidelity will decrease near exponentially with an additional fluctuation when the angle is larger than 0. With a maximum $20\%$ noise inserted in the control, the fidelity decrease $6\%$. Moreover, with a $10\%$ noise inserted in all CZ gates of a 4-qubit quantum full adder, it requires $34$ repetitions to detect the fault. With a $10\%$ noise inserted in one of the CZ gates, the fault turns out to be detected in 100 to 200 repetitions, depending on the gate number. 

This article is organized as follows.  Section II reviews the basics of quantum gates and circuits. In section III, we recall the basic physical model of superconducting qubits. Based on it, in Section IV, we propose our fault model of superconducting quantum circuits. In section V, the simulation tool we employed, QuTiP, is briefly introduced. The last section presents the simulation results. 

\section{Basics of quantum circuits}

The basic data storage element in quantum circuits is quantum bit (qubit, for short). A qubit is a two-level quantum state, which can be expressed as $|\psi\rangle = a|0\rangle + b|1\rangle$, with the probability amplitudes satisfying $|a|^2+|b|^2 = 1$. In general, a state of an $n$-qubit system can be represented as a superposition of the form: $|\psi\rangle = \sum_{i=0}^{2^n-1} c_i |i\rangle$, with integer $i$ being identified with its  binary encoding as an $n$ bit string. An important characteristic of multi-qubit systems is the entanglement between different qubits. This means that one cannot determine the state of each qubit independently even if the qubits are distant from each other. A famous example  of entanglement is the EPR pair, or 2-D Bell state, $\frac{|00\rangle+|11\rangle}{\sqrt{2}}$, which is the maximally entangled state of two qubits. 

Quantum circuit is a basic quantum computing model consisting of quantum gates.  %basic operators of qubits are unitary matrices,. 
A gate operating on an $n$-qubit system is mathematically modelled by a $ 2^n \times 2^n$ unitary matrix, which keeps the norm of state  vectors unchanged. %Meanwhile, quantum circuits consists of a set of unitary gates which operate on qubits included in the circuit. 
%As an example, Fig. \ref{fig:DJ}(a) presents the circuit diagram of Deutsch-Josza algorithm {\color{red}?????}, one of the first quantum algorithms showing quantum speed up over classical computing. 
%which is a typical example of quantum circuits. 
\begin{figure}
    \centering
    \subfloat[\label{fig:a}]{\includegraphics[width = .90\linewidth]{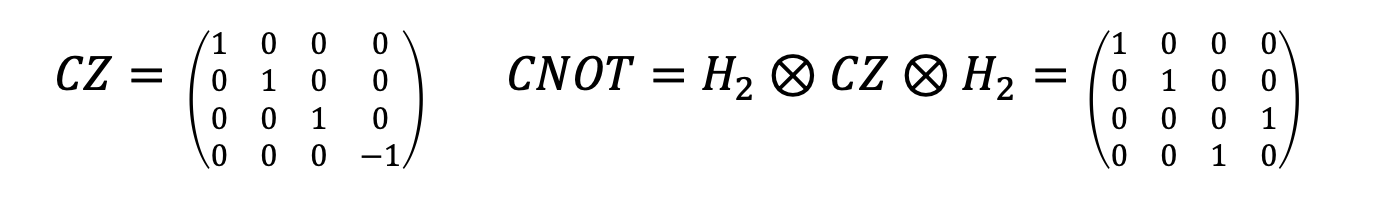}}
    \\
    \subfloat[\label{fig:b}]{\includegraphics[width = .95\linewidth]{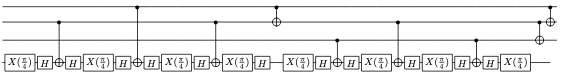}}
    \caption{(a) Matrix representation of CZ and CNOT gates. (b) The quantum circuit for four-qubit quantum full adder. }
    \label{fig:DJ}
\end{figure}
Quantum controlled-Z (CZ) gate is one of the most commonly used two-qubit gates in superconducting quantum hardware, from which the controlled-NOT (CNOT) gate can be generated (see Fig. \ref{fig:a} for the unitary matrices of CZ and CNOT). 
%In superconducting quantum hardware, we have chosen CPhase gate as the elementary gate for a universal gate set. 
In our work, we employ the quantum full adder \cite{Li_2020} and a random circuit as examples in our fault simulation task. They  
consistss of only single qubit gates and CNOT gates. Fig. \ref{fig:b} illustrates the circuit of a 4-qubit quantum full adder.

In general, to include the interaction of qubits with the environment, a noisy quantum gate needs to be described by a quantum channel instead of a unitary operation. Mathematically, the discrete-time description of a quantum channel is a superoperator transforming a mixed quantum state as a density matrix $\rho$ to $\mathcal{E}(\rho) = \sum_{i=1}^{n}E_i\rho E_i^{\dag}$, the set of matrices $\{E_i\}$ is called Kraus operators. The continuous description of a quantum operation is given as the  Liouvelle-von Neumann-Lindblad master equation \cite{2006The}:
%is introduced to describe the quantum operation in a differential equation:
\begin{equation}\label{Lind}
\frac{d\rho}{dt} = -\frac{i}{\hbar}[H,\rho] + \sum_{i=1}^{N^2-1}\gamma_i[L_i\rho L_i^{\dag}-\frac{1}{2}{\rho,L_i^{\dag}L_i}]
\end{equation}
In the master equation, the interaction with environment is described by the collapse operator $L_i$ together with the parameter $\gamma_i$. 

\iffalse
Testing quantum circuits differs from testing traditional logic circuits in that the input and output state are all probabilistic, requiring thorough consideration of all possible output states given a certain input state. Therefore, novel method must be apply to measure the fidelity of real quantum circuits and detect the type and the position of faults. 
\fi

\section{Physical model of coupled superconducting qubits}
Superconducting qubits encode the lowest two energy levels of the nonlinear LC circuits as $|0\rangle$ and $|1\rangle$.  Because of the nonlinearity brought by the Josephson junction, computational and non-computational states are separated. The most widely used design is Transmon \cite{Koch_2007}, a modified charge qubit. To read out the qubit state, a cavity resonator is often coupled to the qubit before connecting it to the bus. 

Since most of the techniques used in realizing the two qubit  controlled-phase (CZ) gate use the avoided level-crossing between $|11\rangle$ and $|20\rangle$, the second excited state $|2\rangle$ should be taken into consideration. The standard circuit quantum electrodynamics (cQED) set-up is two superconducting qubits coupled to a common bus resonator \cite{Magesan_2020}:
\begin{equation}
\begin{aligned}
    H_{\rm{sys}} &= \sum_{j = 1,2}\omega_j(b_j^{\dag}b_j) + \frac{\alpha}{2}(b_j^{\dag}b_j)(b_j^{\dag}b_j - \mathbbm{1}) \\
    &+ \omega_r c^{\dag}c + \sum_{j = 1,2}g_j(b_j^{\dag}c +b_j c^{\dag})
\end{aligned}
\end{equation}
where $b_j$ and $c$ are the destroy operators of qubit $j$ and the resonator respectively, $\omega_j$ is the frequency of $j^{th}$ qubit and $\alpha$ is the anharmonicity, which is set to be consistent for all qubits. However, in the dispersive regime, the detuning between the bus resonator and the connected qubit, $\omega_r - \omega_j$, is much larger than that between a pair of qubits. Therefore, the Hamiltonian of two coupled superconducting qubits can be simplified as  \cite{PhysRevApplied.11.034030}: 
\begin{equation}
    H_{\rm{sys}} = \sum_{j}\hbar \omega_{j}|1\rangle\langle1| + \hbar (2\omega_j + \alpha)|2\rangle\langle2| + \hbar g(J_1^{\dag}J_2+J_1 J_2^{\dag})
\end{equation}
where $J_{1(2)}$ is the annihilation operator for a qutrit system, i.e. $J_{1(2)} = |0\rangle\langle1| + \sqrt{2}|1\rangle\langle2|$.
Here, the rotating wave approximation (RWA) is employed to eliminate the fast rotating terms. We denote the eigenstates of $H_{\rm{sys}}$ and the corresponding energies as $\{|\overline{ij}\rangle, E_{ij}\}$.

\section{Fault models of superconducting quantum circuits}

In this section, we present our fault models for superconducting quantum circuits, which can serve as a basis for various EDA tasks for quantum computing like fault simulation and ATPG. 

We focus on the faults in the controlled-Z (CZ) gate since two qubit gates usually have lower gate fidelities than single qubit gates in real superconducting quantum hardware. In the past decades, three approaches to realizing the CZ gate have been developed: (i) cross resonance for fixed frequency qubits \cite{PhysRevB.81.134507} \cite{Allen_2017} \cite{Tripathi_2019}, (ii) adiabatic \cite{PhysRevApplied.11.034030} \cite{DiCarlo_2009} \cite{PhysRevA.87.022309} \cite{Martinis_2014}
 or diabatic \cite{barends2019diabatic} frequency tuning for frequency tunable qubits, and (iii) frequency modulation of tunable coupler \cite{PhysRevA.97.022330} \cite{PhysRevB.87.220505} \cite{PhysRevApplied.10.034050}. %Since most of the superconducting quantum hardware consist of frequency tunable qubits and use frequency tuning strategy to realize CZ gates
We will mainly consider the frequency tuning approach. 
\subsection{Frequency Modulating Approach}
\begin{figure}
    \centering
   
   \includegraphics[width = .90\linewidth]{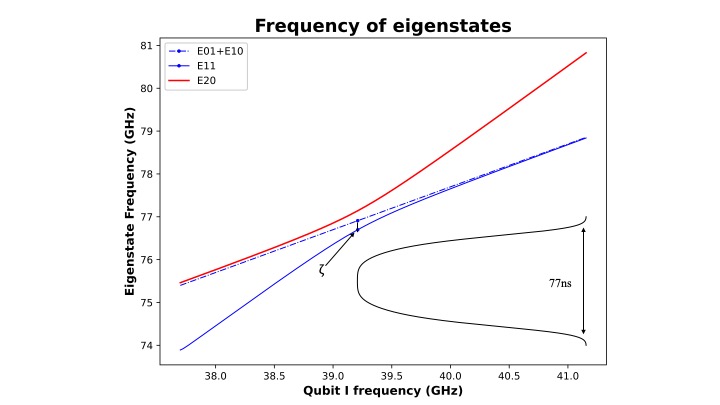}
    \caption{Energies of eigenstates $E_{10}+E_{01}, E_{11}, E_{20}$. The avoided level-crossing between $|11\rangle$ and $|20\rangle$ and the static ZZ coupling strength $\zeta$ is shown in the figure. The pulse with gate duration 77ns lies in the lower right corner.}
    \label{fig:czgate}
\end{figure}

Adiabatic frequency tuning approach modulates the frequency of the control qubit to harness the avoided level-crossing effect between $|11\rangle$ and $|20\rangle$ to activate the strong $\sigma_z\otimes \sigma_z$ interaction. Fig. \ref{fig:czgate} shows the energy shift  and a typical frequency tuning trajectory \textemdash \ Slepian.
The unitary propagator induced by the frequency modulating is:
 \begin{align}
    \setlength{\arraycolsep}{1.2pt}
    U = \begin{array}{lc}
    \begin{pmatrix}
    1 & 0 & 0 & 0  \\
    0 & e^{-i\phi_{01}} & 0 & 0 \\
    0 & 0 & e^{-i\phi_{10}} & 0 \\
    0 & 0 & 0 & e^{-i\phi_{11}} \\
    \end{pmatrix}
    \end{array}
    \end{align}
where the relative phases $\phi_{ij} \approx \int_{0}^{t_{\rm{gate}}}E_{ij}(t) dt$ and $E_{ij}(t)$ is the energy of the eigenstates $|\overline{ij}\rangle$ from the dynamical Hamiltonian $H_{\rm{sys}}(t)$. Because of the avoided level-crossing, the state $|11\rangle$ will accumulate an additional phase $\varphi$, that is, $\phi_{11} = \phi_{01}+\phi_{10} + \varphi$. Therefore,  performing a CZ gate is to set $\varphi \approx (2n+1)\pi$ where n is an integer. We mainly apply the quasi-adiabatic approach from \cite{Martinis_2014} and inverse-Gaussian waveform from \cite{PhysRevA.87.022309}, in which the frequency changes rapidly when far from detuning and evolves adiabatically near resonance.

\subsection{Fault models}

Faults in superconducting quantum circuits can be divided into two categories: coherent faults 
%(which can be described by unitary matrices) 
and incoherent faults %(which are represented by superoperators)
. In  the frequency modulating approach, we consider two kinds of coherent faults and two kinds of incoherent faults: 

\textbf{Coherent faults:} Coherent faults can be described by unitary fault matrices, and can be corrected according to simulation results provided by fault models. Our fault models mainly deal with the following coherent faults: 
\begin{itemize}
    \item \textbf{Pulse faults:} 
        Pulse faults mainly arise from the errors in control parameters and gate duration. In  the framework of frequency control, the additional dynamical phase is accumulated by $\varphi(t) = \int_0^{t_{\rm{gate}}}( E_{11}'(t) - E_{11}(t)) dt$. 
        Considering the subspace spanned by $\{|11\rangle, |02\rangle\}$, the dynamical phase is expressed as \cite{PhysRevApplied.11.034030}: 
      \begin{equation}
        \varphi(t) = \int_0^{t_{\rm{gate}}}\sqrt{2}g(\tan(\frac{\sqrt{2}g}{\omega_1(t)- \omega_2+\alpha}))dt
        \end{equation}
    In this work, we apply one of the nearly optimal frequency waveforms -- Slepian function \cite{Martinis_2014} as the differential of the waveform to minimize the power spectral density. Because the Slepian function has no analytical form, we use its Fourier approximation:
    \begin{equation}
      \frac{d\theta}{dt} = \rm{sgn}(t - \frac{t_{\rm{gate}}}{2})\sum_{n = 1,2,\cdots,m}\lambda_n[1-\cos(2\pi nt/t_{\rm{gate}})]
    \end{equation}
    where $\theta = \frac{\sqrt{2}g}{\omega_1(t)- \omega_2+\alpha}$, $m$ is preset and the parameters $\{\lambda_n\}$ are determined by interpolation.
    
    \ \ \ The faults come from the errors in both waveform parameters and gate duration. We mainly consider ratio and bias error for waveform parameters, and truncation error for gate duration. Therefore, suppose we use $m$ Fourier terms to approximate Slepian function and there are $n$ CZ gates in the quantum circuit, the number of pulse faults will be \textbf{$(2m+1)n$}.
    \item \textbf{Missing gate faults:} 
    The missing gate fault describes the error when a gate is not executed. Here we use a more realistic model. Instead of a trivial identity operation, a unitary generated by the Hamilton with no control pulse is used to include the intrinsic error of the system. Even if the control pulse is missing, there is still a remaining ZZ coupling between the two qubits, therefore the missing gate fault is also called idling fault. Consequently, this static ZZ interaction becomes the dominating term of system dynamics. 
    
    Therefore, the faulty gate \textemdash \  an undesired rotation induced by the missing gate fault \textemdash  \ should be:
    \begin{equation}
       U_{F} \approx \rm{diag}(1,1,1,e^{i \zeta t})
    \end{equation}
 where $\zeta$ represents the static ZZ shift strength introduced by the coupling term. The static ZZ shift strength can be either expressed as $\zeta = \omega_{11} - \omega_{10} - \omega_{01}$ or be measured using a Ramsey-type experiment \cite{PhysRevResearch.2.033447}. Moreover, the period of idling is $T= \frac{2\pi}{\zeta}$. Consequently, the missing gate fault co So in a quantum circuit with $n$ CZ gates after decomposing, there are totally $n$ possible missing gate faults. 
\end{itemize}

\textbf{Incoherent faults:}
\iffalse
Incoherent faults considered here are system noises that cannot be described by unitary matrices. They cannot be corrected simply by adjusting the controls according to the fault models, and have already go beyond the scope of current fault-tolerant quantum computation. Therefore, benchmarking methods like Randomized Benchmarking \cite{Knill_2008} should first be applied to characterize those incoherent faults. In this part we discuss two important incoherent faults in superconducting quantum circuits: 
\fi
\begin{itemize}
 \item \textbf{State leakage faults:} Although the generalized leakage faults may be suppressed by improving pulse schemes, for simplicity, we assume here that, except for the $|11\rangle-|20\rangle$ transition, the leakage population is lost and never returns back to the computational states. Under such an assumption, leakage fault will act as an incoherent fault, causing the fidelity to decrease. The overall state leakage effects can be described by a noise generator of the CZ gate \cite{Ghosh_2013}. 
 
    \begin{align}
    \setlength{\arraycolsep}{0.8pt}
    S' = \begin{array}{lc}
    \begin{bmatrix}
    0 & 0 & 0 & 0& 0 & 0  \\
    0 & \zeta_1 & 0 & i\chi_1 e^{i\phi_1}& 0 & 0 \\
    0 & 0 & 0 & 0& i\chi_2 e^{i\phi_2} & 0 \\
    0 & -i\chi_1 e^{-i\phi_1} & 0 & \zeta_2 & 0 & 0 \\
    0 & 0 & -i\chi_2 e^{-i\phi_2} & 0& \zeta_3 & i\chi_3 e^{i\phi_3}   \\
    0 & 0 & 0 & 0 & -i \chi_3 e^{-i\phi_3} & \zeta_4 \\
    \end{bmatrix}
    \end{array}
    \end{align}

    \medskip

    The rows and columns of $S'$ is
    $|00\rangle,|01\rangle,|02\rangle,|10\rangle,|11\rangle,|20\rangle$. Then the noisy operator of CZ gate can be expressed by $U' = Ue^{iS'}$. The decrease of fidelity due to state leakage should be $\Delta F = 1/d^2 \rm{tr}|e^{iS'}|^2$, where $d$ is the dimension of $S'$. In real physical devices the parameters are set to $\chi_i \approx 10^{-2}, \zeta_i \approx 10^{-2}$.

    \item \textbf{Decoherence:} As for superconducting quantum hardware, relatively short coherent time is a critical drawback, which means the entanglement of qubits is gradually lost due to the coupling with the environment. To describe the effect of decoherence, we use the Liouvelle-von Neumann-Lindblad master equation (\ref{Lind}). 
    %\cite{2006The}: 
    %\begin{equation}
        %\frac{d\rho}{dt} = -\frac{i}{\hbar}[H,\rho] + \sum_{i=1}^{N^2-1}\gamma_i[L_i\rho L_i^{\dag}-\frac{1}{2}{\rho,L_i^{\dag}L_i}]
   % \end{equation}
    For a 3-level qutrit system, the Lindblad operators for decoherence is $L_1 =|0\rangle\langle 1| + \sqrt{2}|1\rangle\langle 2|$, $L_2 = |1\rangle\langle 1| + 2|2\rangle \langle 2|$, which represents the amplitude damping and pure dephasing effects,  respectively.
    There are mainly two kinds of decoherence effects:
    \begin{itemize}
        \item \textit{Amplitude damping (energy relaxation):} When amplitude damping occurs, the population of excited states will decrease, which means the $|1\rangle$ component of the quantum state will slowly evolve to $|0\rangle$. Therefore, amplitude damping also indicates the energy relaxation of the system. 
        \iffalse
        The amplitude damping effect of a three-level quantum state could be described by the following super operators: 
        \begin{equation*}
        \begin{aligned}
            E_1 = \begin{bmatrix}
            1 & 0 & 0 \\
            0 & \sqrt{1 - \lambda_1} & 0 \\
            0 & 0 & \sqrt{1 - \lambda_2}
            \end{bmatrix} 
            , \\ E_2 = \begin{bmatrix}
            0 & \sqrt{\lambda_1} & 0 \\
            0 & 0 & 0 \\
            0 & 0 & 0
            \end{bmatrix}, \quad
            E_3 = \begin{bmatrix}
            0 & 0 & \sqrt{\lambda_2} \\
            0 & 0 & 0 \\
            0 & 0 & 0
            \end{bmatrix}
        \end{aligned}
        \end{equation*}
        where $\lambda_m = 1 - e^ {-m\Delta t/T_1}$, $T_1$ is the characteristic time of energy relaxation. 
        \fi
        To estimate the charge qubit's (or Transmon's) characteristic time of energy relaxation $T_1$, Purcell effect induced by the dispersive cavity resonator needs to be considered \cite{Koch_2007}. The altered relaxation rate induced by Purcell effect can be described as: 
        \begin{equation}
            \Gamma_k^{i,i+1} = \kappa \frac{g_{i,i+1}^2}{(\omega_{i,i+1} - \omega_r)^2}
        \end{equation}
        In real experiments we often take the lifetime of the resonator as $1/\kappa = 160ns$, while the coupling strength and anharmonicity between resonator and qubit frequency are given by the parameters of devices. 
        
        \item \textit{Phase damping (dephasing)}: When phase damping occurs, the projection of the state vector onto the X-Y plane of a Bloch sphere will slowly evolve to zero, indicating that the phase information of the state vector is gradually dropped out. If there only exists the phase damping effect, then only those off-diagonal elements of the density operator will decay at the rate $T_\phi$ without affecting diagonal elements. 
        
       \ \ \ Same as the relaxation rate $\Gamma_1 = 1/T_1$, the pure dephasing time $T_\phi$ also comes from various physical effects. For Transmon-based qubits, the dominating effect for $T_\phi$ is the critical current noise \cite{Koch_2007} rather than the charge noise in traditional cooper pair box (CPB) devices. Since the special design of Transmon makes charge noise decrease exponentially with the ratio of $E_J/E_C$, Transmon based qubits are insensitive to charge noise. In real cases, a Transmon device with $ A = 10^{-6}I_c$ and $E_J = 35$ GHz, $E_C = 0.4$ GHz will have the pure dephasing time $T_{\phi} \approx 52\mu s.$ 
    \end{itemize}
The total dephasing time $T_2$ is expressed as $\frac{1}{T_2} = \frac{1}{2T_1}+\frac{1}{T_\phi}$. In real superconducting quantum hardware, $T_2$ is usually $20-50\mu s$. \cite{PhysRevResearch.2.033447} \cite{Koch_2007}. 
\end{itemize}
\subsection{A general workflow for using fault models}
\begin{figure}
    \centering
    \includegraphics[width = .85\linewidth]{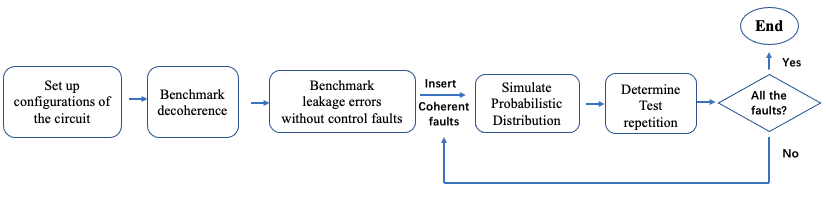}
    \caption{A general workflow when simulating effects of fault models in superconducting quantum circuits.}
    \label{fig:workflow}
\end{figure}

We propose a workflow for using these fault models defined above and summarized in Table \ref{tab:faultmodel}. When conducting EDA tasks in superconducting quantum circuits, the benchmark results including only decoherence is first provided by a master equation solver. Then faults induced by leakage are considered, also as a benchmark for real quantum hardware. After that we insert noises in control coefficients and conduct fault simulation on quantum circuits and find the minimal test repetitions for each fault and fault coverage for a given test pattern. This workflow is visualized in Fig. \ref{fig:workflow}. 

\begin{table*}[]
    \centering
    \begin{tabular}{c|c|c|c}
    \toprule
        Faults & Category &Physical Factors &Fault effect\\
        \midrule
        Pulse faults & Coherent & Noise in control parameters & Undesired ZZ rotation \\
        Leakage faults & Incoherent & Non-adiabaticity in pulse & Population decrease of computational basis \\
        Decoherence & Incoherent & Superconducting system & off-diagonal element dissipation \\
        Missing gate faults & Coherent & Static ZZ shift & Undesired ZZ rotation \\
    \bottomrule
    \end{tabular}
    \caption{A brief summary to fault models in superconducting quantum circuits.}
    \label{tab:faultmodel}
\end{table*}

\section{Simulation tool -- QuTiP} Now we turn to the fault simulation of superconducting quantum circuits based on the fault models developed in the last section. 
We choose to use QuTiP \cite{JOHANSSON20131234} -- an  open-source toolbox for quantum computing in Python -- to carry out our simulation tasks. QuTiP provides efficient solvers for time-dependent noisy Hamiltonians such as \textit{mesolve} for Lindblad master equations, which represent quantum systems coupled with external environment. QuTiP also helps convert the gate-level simulation in real quantum hardware to pulse-level simulation tasks  \cite{li2021pulselevel}. 
\iffalse
In the next section, we will use the new toolbox \texttt{qutip-qip} \cite{li2021pulselevel} to perform quantum circuit simulation with fault models added to the pulse. 
\fi

\section{Experiments}
\subsection{Evaluation metrics and Simulation design}
We choose the fidelity of two density operators as our metric for noisy quantum circuits. The fidelity is defined as \cite{Raginsky_2001}: 
\begin{equation}
    \mathcal{F}(\rho,\sigma) = tr(\sqrt{\sqrt{\rho}\sigma\sqrt{\rho}})^2
\end{equation}
 A counterpart of density operator fidelity  can also be defined for noisy quantum gates \cite{Raginsky_2001} : \begin{equation}
    \mathcal{F}(U_{\rm{ideal}},U_{\rm{real}}) = \frac{1}{d^2}|tr(U_{\rm{ideal}}^{\dag}U_{\rm{real}})|^2
\end{equation} where $d$ is the dimension of the Hilbert space on which these gates act. Both of them will be used to evaluate the effect of noise on quantum gates or circuits. 

We first calibrate the pulse of the CZ gate for the maximal fidelity without decoherence and then investigate the effect of coherent faults. After that we insert coherent faults into CZ gates and carry out fault simulation on two 4-qubit quantum circuits to obtain the test repetitions. The maximum fidelity benchmark of different pulse shapes is shown in Table \ref{tab:bench}, in which the pulse shape named after 'Hanning window', 'Fourier-2', 'Fourier-4' are Fourier approximations of Slepian function by using one, two and four Fourier terms each.  

\begin{table}
\centering
  \caption{Simulation results of simple control pulses. The accuracy of the gate time is limited to 0.1ns.}
  \label{tab:bench}
  \begin{tabular}{ccccl}
    \toprule
    Pulse shape & Max Fidelity & Pulse shape & Max Fidelity\\
    \midrule
    Square  &  $99.30\%$ & Hanning window  & $99.93\%$ \\
    Cosine  &  $99.90\%$ & Fourier-2 & $99.95\%$ \\
    Slepian &  $99.99\%$ & Fourier-4 & $99.98\%$ \\
    
  \bottomrule
\end{tabular}
\end{table}
\subsection{Simulation in Single gate}
\subsubsection{Missing gate faults}
We first consider fault simulation of missing gate faults. By transforming the idling Hamiltonian into the rotating frame, it can be seen that the variation of fidelity can be described by a rapid controlled-Z rotation in the two-qubit system. The missing gate fidelity at the maximum gate duration 77.0ns is 0.802. Simulation results of missing gate faults in single CZ gate show that whichever the CZ gate, simulating the circuit once suffices to detect the missing gate fault with 99\% confidence level. 
\iffalse
  \begin{figure}
        \centering
        \includegraphics[width = 0.8\linewidth]{Figures/idling.png}
        \caption{Illustration of the effect of missing gate fault, in the computational level $|10\rangle$ and $|11\rangle$ of two-qubit gates. With the existence of coupling, the system state goes through an oscillation at the period $\frac{2\pi}{E_2' - E_1'}$.}
        \label{fig:missingate}
        \end{figure}
\fi
\subsubsection{Control pulse error}
\begin{figure}
     \centering
    \subfloat[\label{fig:fa1}]{\includegraphics[width = .45\linewidth]{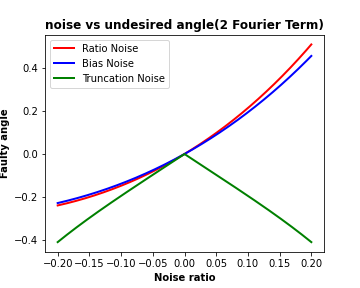}}
    \subfloat[\label{fig:fa2}]{\includegraphics[width = 
    .45\linewidth]{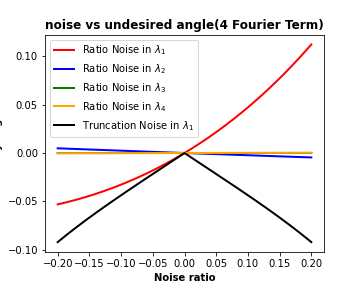}}
    \caption{(a) Undesired rotation angle brought by ratio, bias and truncation noise in $\lambda_1$ of 'Fourier-2' waveform. (b) Undesired rotation angle brought by ratio noise in each of the coefficient of 'Fourier-4' waveform and truncation noise in $\lambda_1$. Both figures illustrate the theoretical estimation without numerical simulation.}
    \label{fig:FAngle}
\end{figure}

To test the fault effect of noises in the control coefficients, we employ the Fourier approximation of the Slepian function, shown in Eq.(\ref{Eq:Fourier2}) and add ratio, bias and truncation noises to the coefficients $\lambda_{n}$. 

\begin{equation}
	\frac{d\theta}{dt} = 
\mathrm{sgn}(t - \frac{t_{\text{gate}}}{2})\sum_{n = 1}^{m}\lambda_n[1-\cos(2\pi n t/t_{\text{gate}})]
\label{Eq:Fourier2}
\end{equation}
\iffalse
Fig. \ref{fig:slepianfourier} illustrates the Fourier approximation of Slepian function, in which four Fourier elements and only one element are used each. 
%%Here the parameters used in the Fourier approximation needs to be updated but haven't done yet. 
\begin{figure}
    \centering
    \includegraphics[width = .80\linewidth]{Figures/Slepian_comp.png}
    \caption{Fourier approximations of Slepian functions, using 4 elements or one element each. }
    \label{fig:slepianfourier}
\end{figure}
\fi
We use the gate duration at the optimal gate fidelity point and then insert ratio, bias and truncation faults into the parameters. We then compute the theoretical undesired rotation angle and the fidelity of the noisy gate. Fig. \ref{fig:FAngle} illustrates the undesired rotation angle versus the noise ratio. It indicates that truncation noise brings a linear undesired angle to the gate while the angle brought by ratio and bias noise is approximately quadratic. Moreover, since $\lambda_1$ is the dominating coefficient in the pulse, the noise in $\lambda_{n},n > 1$ could be ignored while the noise in $\lambda_1$ is present.

\begin{figure}
    \centering
   \subfloat[\label{fig:ff1}]{\includegraphics[width = .45\linewidth]{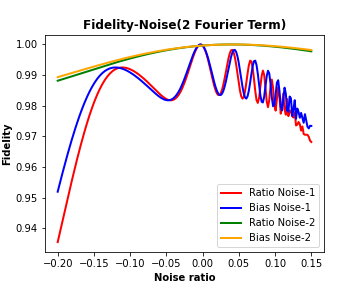}}
    \subfloat[\label{fig:ff2}]{\includegraphics[width = 
    .45\linewidth]{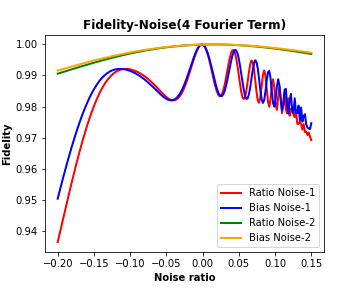}}
    \caption{(a) Fidelity versus noise in the Fourier waveform with 2 terms. (b) Fidelity versus noise in the Fourier waveform with 4 terms. Both figures illustrate the real result simulated by master equation.}
    \label{fig:Ffid}
\end{figure}
Fig. \ref{fig:Ffid} shows the fidelity versus the noise ratio obtained by real results simulated by master equation. Slightly different from the theoretical results illustrated in Fig. \ref{fig:FAngle}, The experimental fidelity goes through more oscillations, which are very common indications for off-resonant leakage-like dynamics. 
\iffalse
\subsection{State leakage error}
To test the state leakage errors, we directly combine the noise generator matrix with the original generator matrix and observe the fidelity decrease due to the state leakage error. Fig. \ref{fig:stateleak} illustrates that the fidelity decrease follows a quadratic relationship with the leakage factor $\chi_i$, which corresponds with what the theoretical noise generator implies. Therefore a $10^{-2}$ error rate of $\chi_i$ will lead to a error of  $10^{-4}$ scale. We can also conclude from the result that the dynamical phases $\phi_i$ will not affect the fidelity. 
\begin{figure}
    \centering
    \includegraphics[width = .90\linewidth]{Figures/result1.png}
    \caption{Fidelity decrease according to the value of leakage factor $\chi_i$. }
    \label{fig:stateleak}
\end{figure}
\fi

\subsubsection{Decoherence}
\begin{figure}
    \centering
    \includegraphics[width = .80\linewidth]{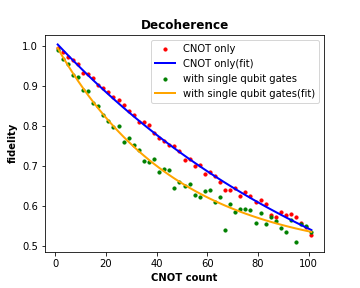}
    \caption{Fidelity change brought by decoherence. The experimental data are shown together with the fitted curves.}
    \label{fig:decoherence}
\end{figure}

In this part we evaluate the effect of decoherence. We set the energy relaxation time $T_1 = 100\mu s$ and total dephasing time $T_{2}^{*} = 20\mu s$. We employ a two-qubit random circuit with $m$ layers to benchmark the decoherence effect. In each layer there are two single qubit gates randomly selected from the set  \{$R_x(\pi/4),R_y(\pi/4),R_z(\pi/4)$\} together with a CNOT gate with a random control qubit. We evaluate the gate fidelity with decoherence in this circuit with depth ranging from $1$ to $101$, and try to fit exponential curves using the experimental data. The experimental and fitting results are shown in Fig. \ref{fig:decoherence}.

\subsection{Simulation in multi-qubit circuits}
In this section we conduct fault simulation in multi-qubit circuits and consider the minimal test repetitions for a given fault. For each test pattern we generate a fault-free output distribution and a faulty one. Then we generate observations by randomly sampling from the faulty output distribution and perform multivariate hypothesis testing by using chi-square values. Under null hypothesis, the chi-square value between the faulty and fault-free output should follow the chi-square distribution:
\begin{equation}
    \sum_{j = 1}^{J}\frac{(n_j - \mu_j)^2}{\mu_j} \sim \chi_{J-1}^2
\end{equation}

\begin{figure}
    \centering
    \includegraphics[width = .90\linewidth]{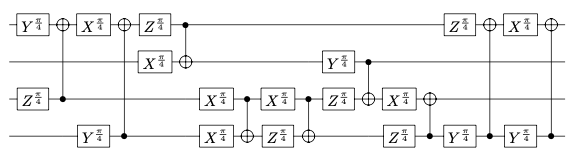}
    \caption{The 4-qubit random circuit with 9 CZ gates for fault simulation.}
    \label{fig:randcir}
\end{figure}
\begin{figure}
    \centering
    \subfloat[\label{fig:rep5}Test repetition for $5\%$ noise]{\includegraphics[width = .45\linewidth]{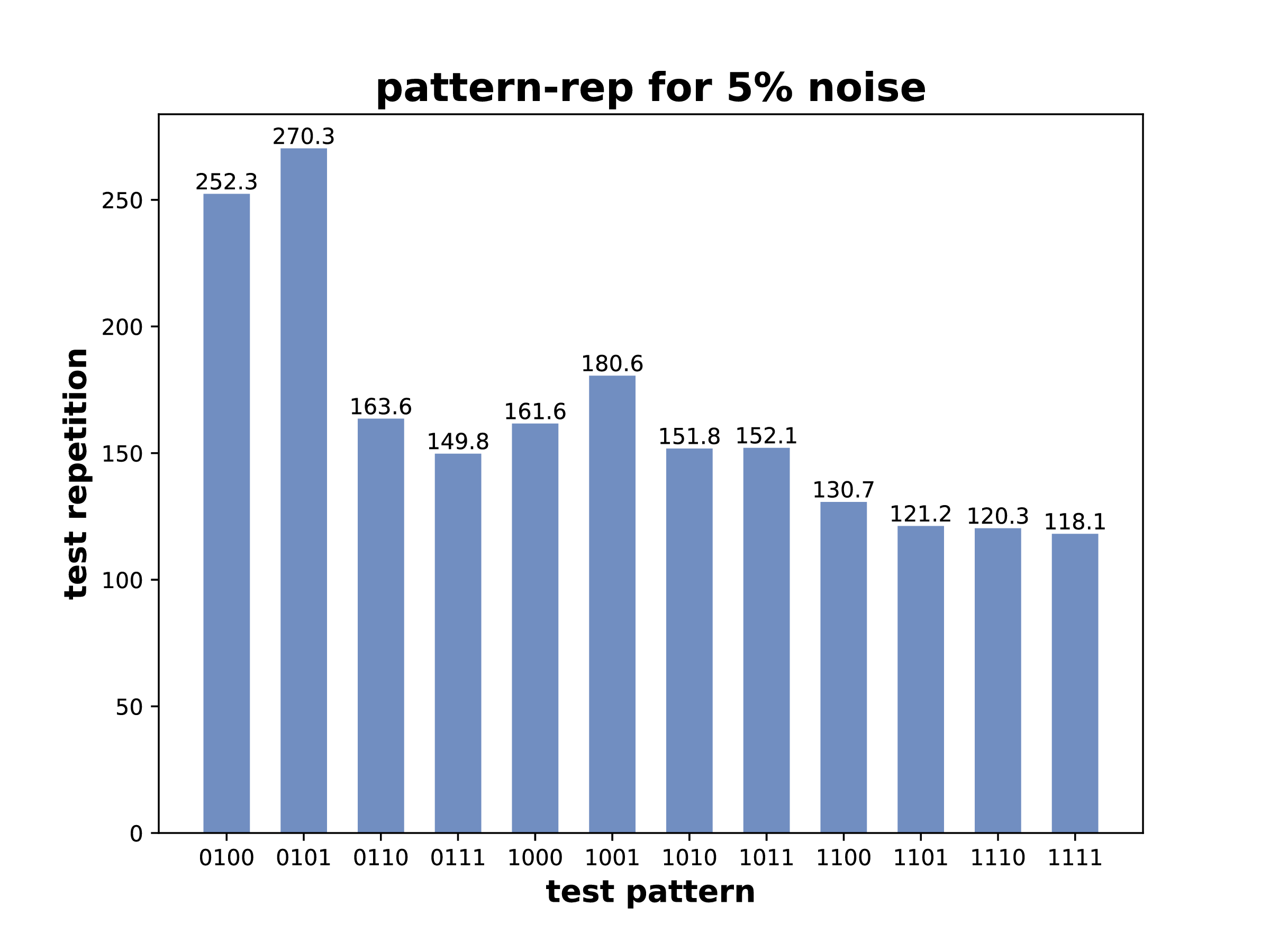}}
    \subfloat[\label{fig:rep10}Test repetition for $10\%$ noise]{\includegraphics[width = .45\linewidth]{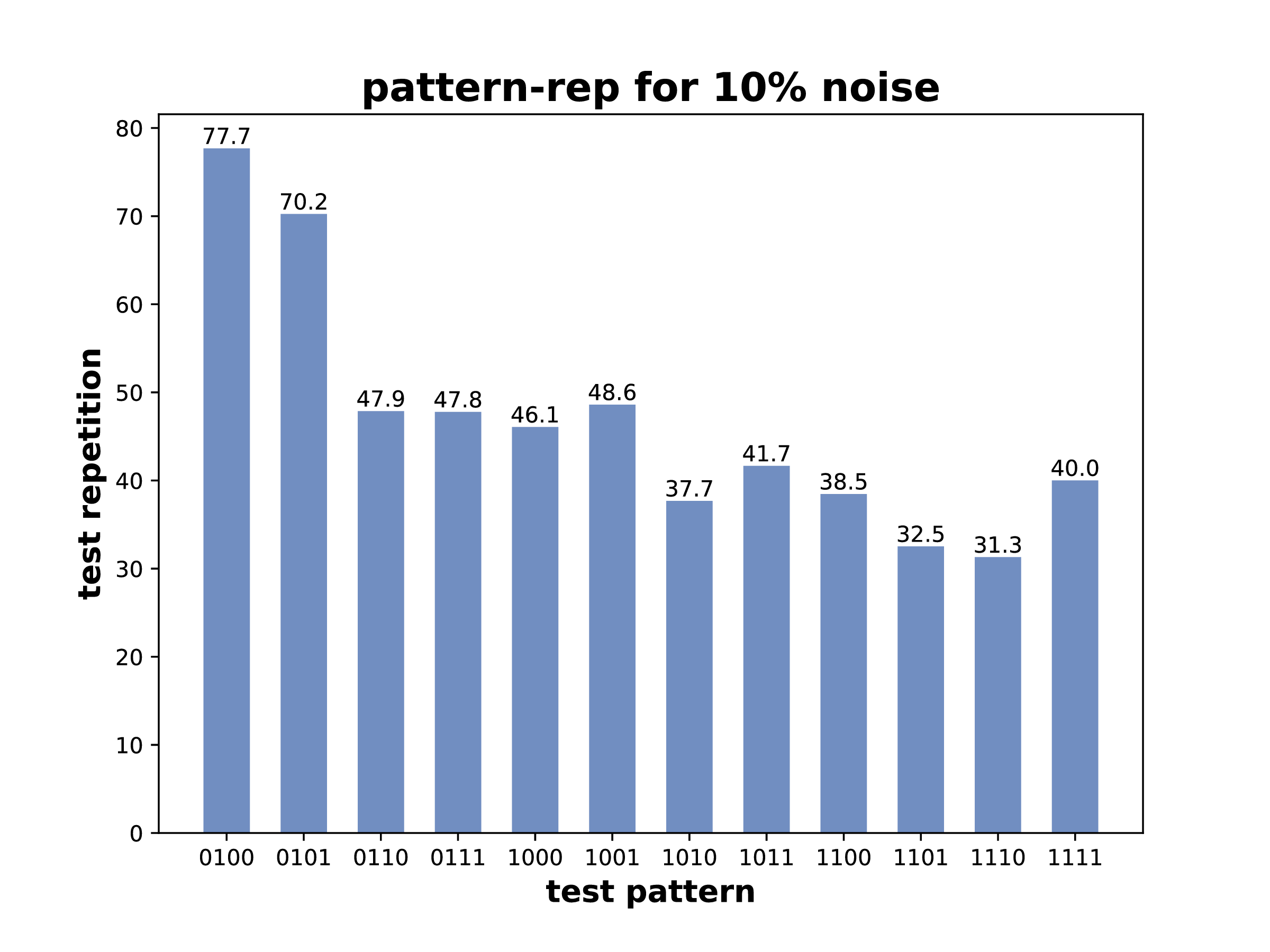}}
    \\
    \subfloat[\label{fig:rep15}Test repetition for $15\%$ noise]{\includegraphics[width = .45\linewidth]{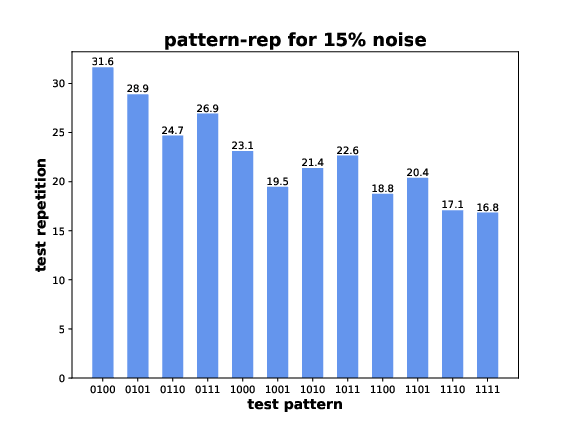}}
    \subfloat[\label{fig:rep20}Test repetition for $20\%$ noise]{\includegraphics[width = .45\linewidth]{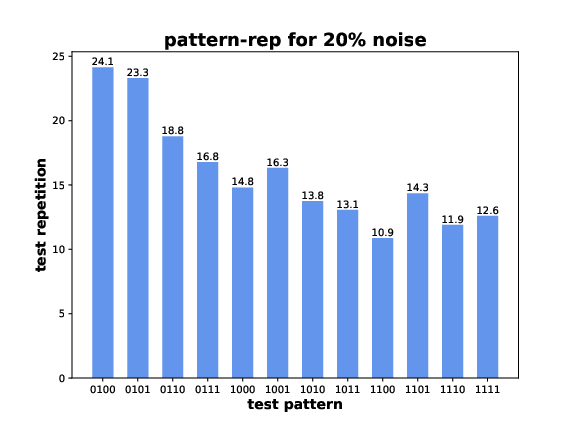}}
    \\
    \caption{Test repetitions for different levels of errors inserted in all of the CZ gates of 4-qubit quantum full adder. The results are averaged over 50 times of sampling.}
    \label{fig:pattern1}
\end{figure}
We set the critical value as 0.99.
Since the chi-square test is carried out by random sampling, for each test pattern we repeat the experiment for 50 times and use the average number of samples as the test repetition. 

\begin{figure}
    \centering
    \subfloat[\label{fig:rep5-rand}Test repetition for $5\%$ noise]{\includegraphics[width = .45\linewidth]{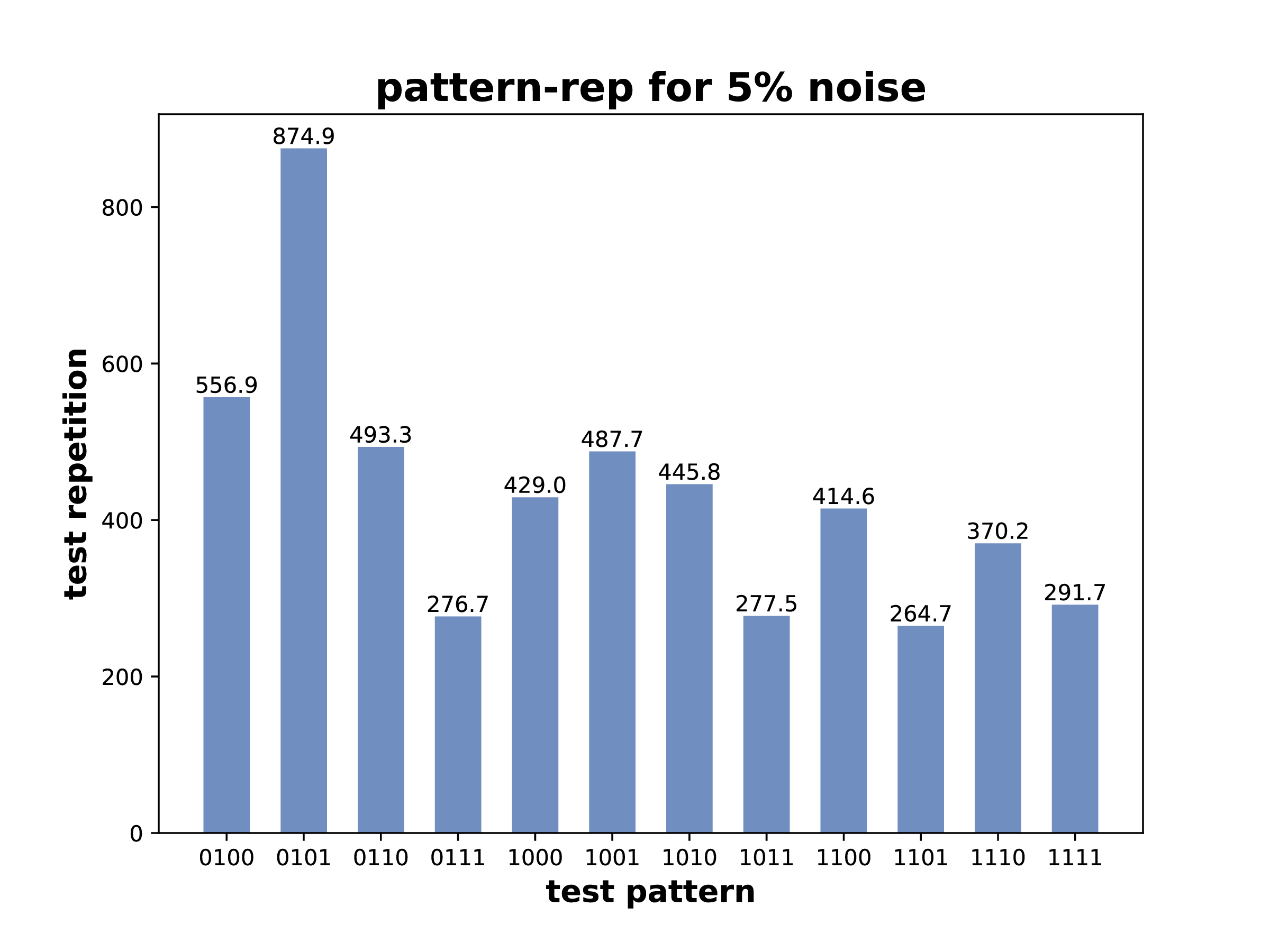}}
    \subfloat[\label{fig:rep10-rand}Test repetition for $10\%$ noise]{\includegraphics[width = .45\linewidth]{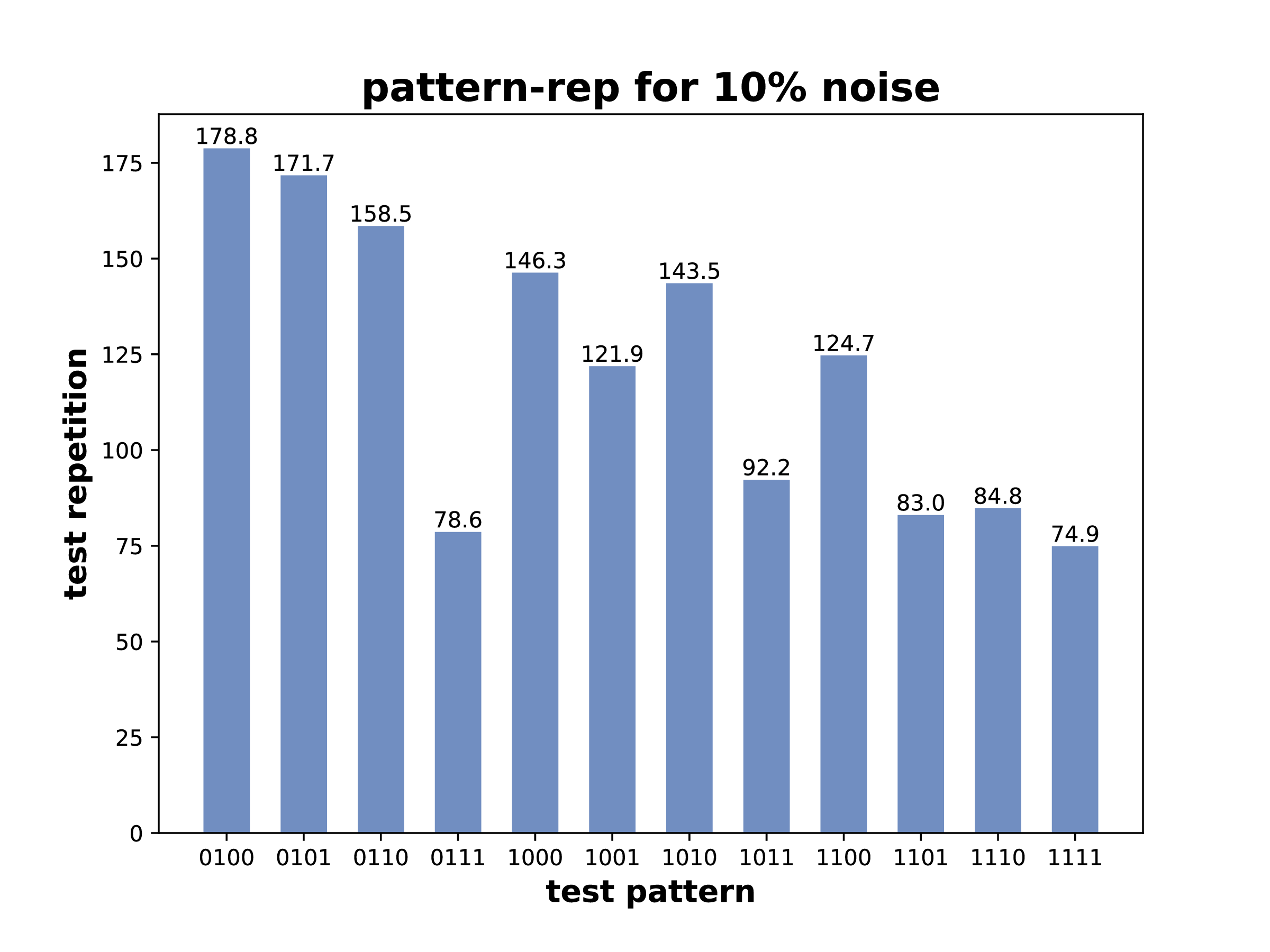}}
    \\
    \subfloat[\label{fig:rep15-rand}Test repetition for $15\%$ noise]{\includegraphics[width = .45\linewidth]{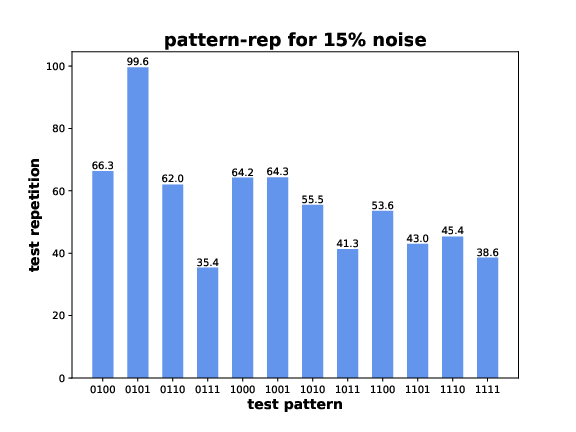}}
    \subfloat[\label{fig:rep20-rand}Test repetition for $20\%$ noise]{\includegraphics[width = .45\linewidth]{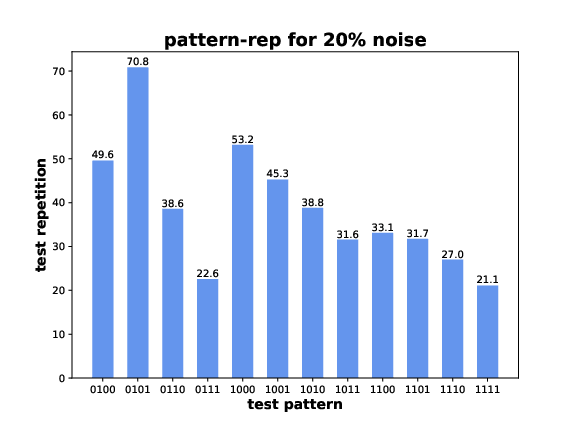}}
    \\
    \caption{Test repetitions for different levels of errors inserted in all of the CZ gates of 4-qubit random circuit. The results are averaged over 50 times of sampling.}
    \label{fig:pattern1-rand}
\end{figure}

\begin{figure}
    \centering
    \subfloat[\label{fig:rep10-ra}Test repetition for $10\%$ ratio noise]{\includegraphics[width = .45\linewidth]{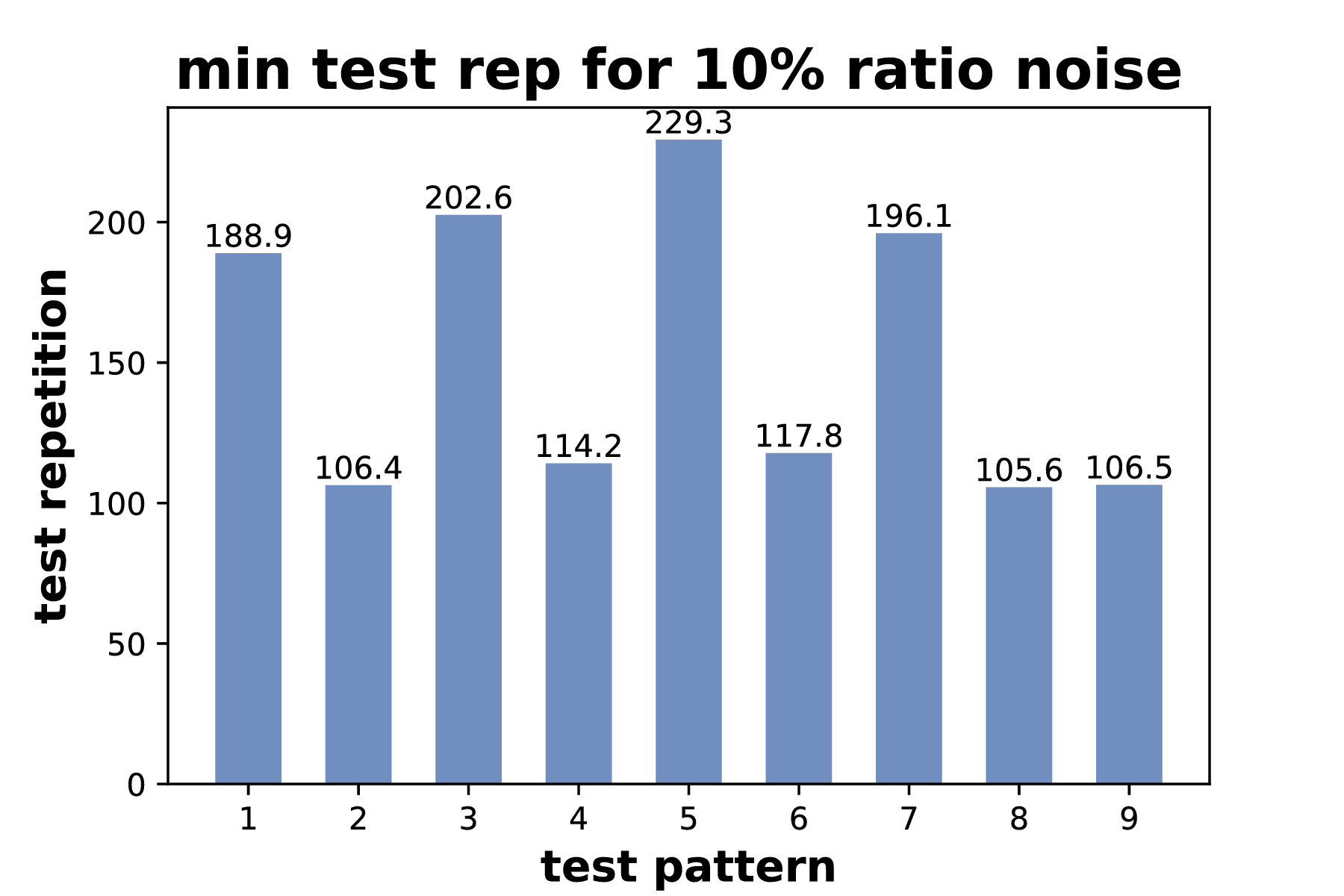}}
    \subfloat[\label{fig:rep20-ra}Test repetition for $20\%$ ratio noise]{\includegraphics[width = .45\linewidth]{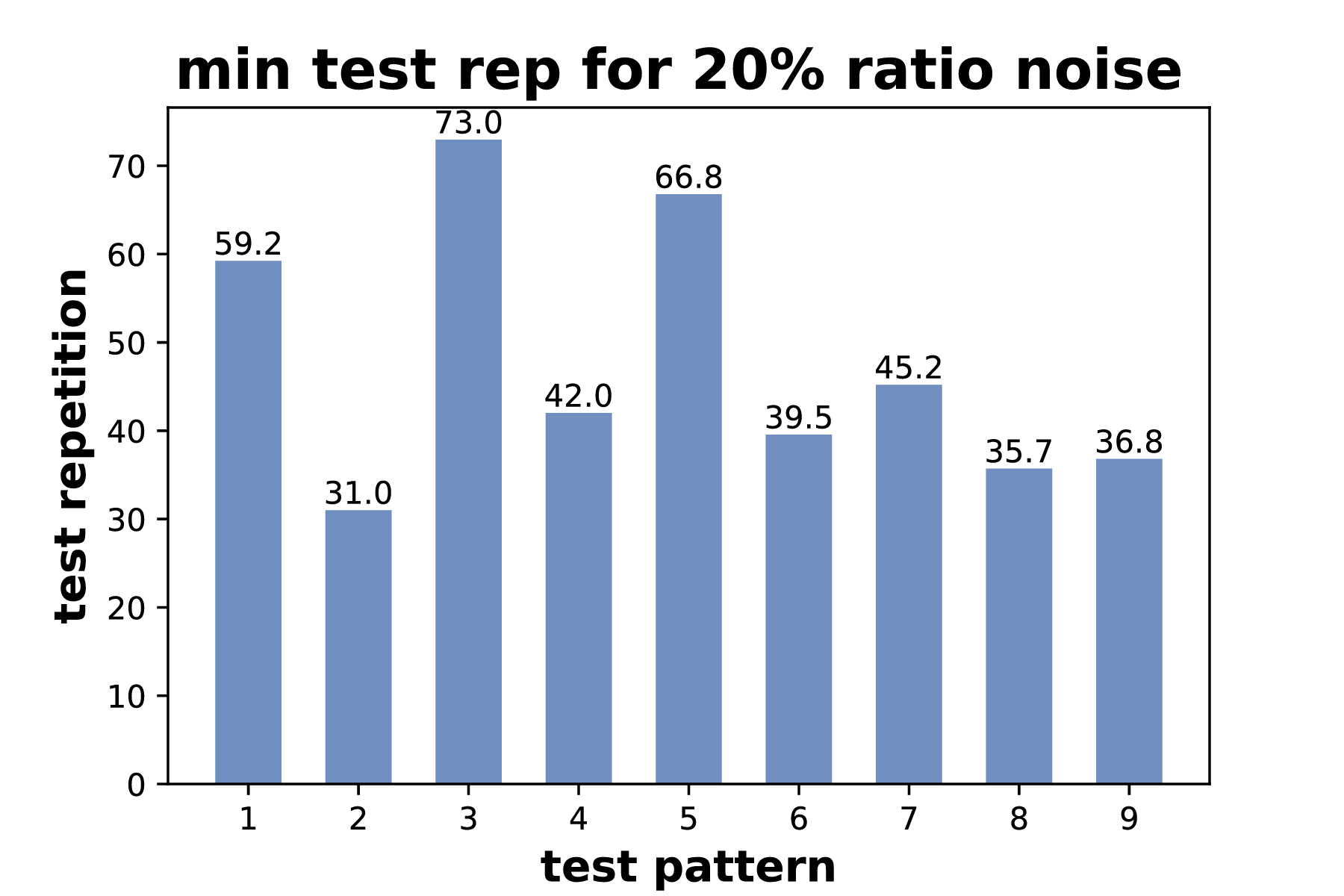}}
    \\
    \subfloat[\label{fig:rep10-bi}Test repetition for $10\%$ bias noise]{\includegraphics[width = .45\linewidth]{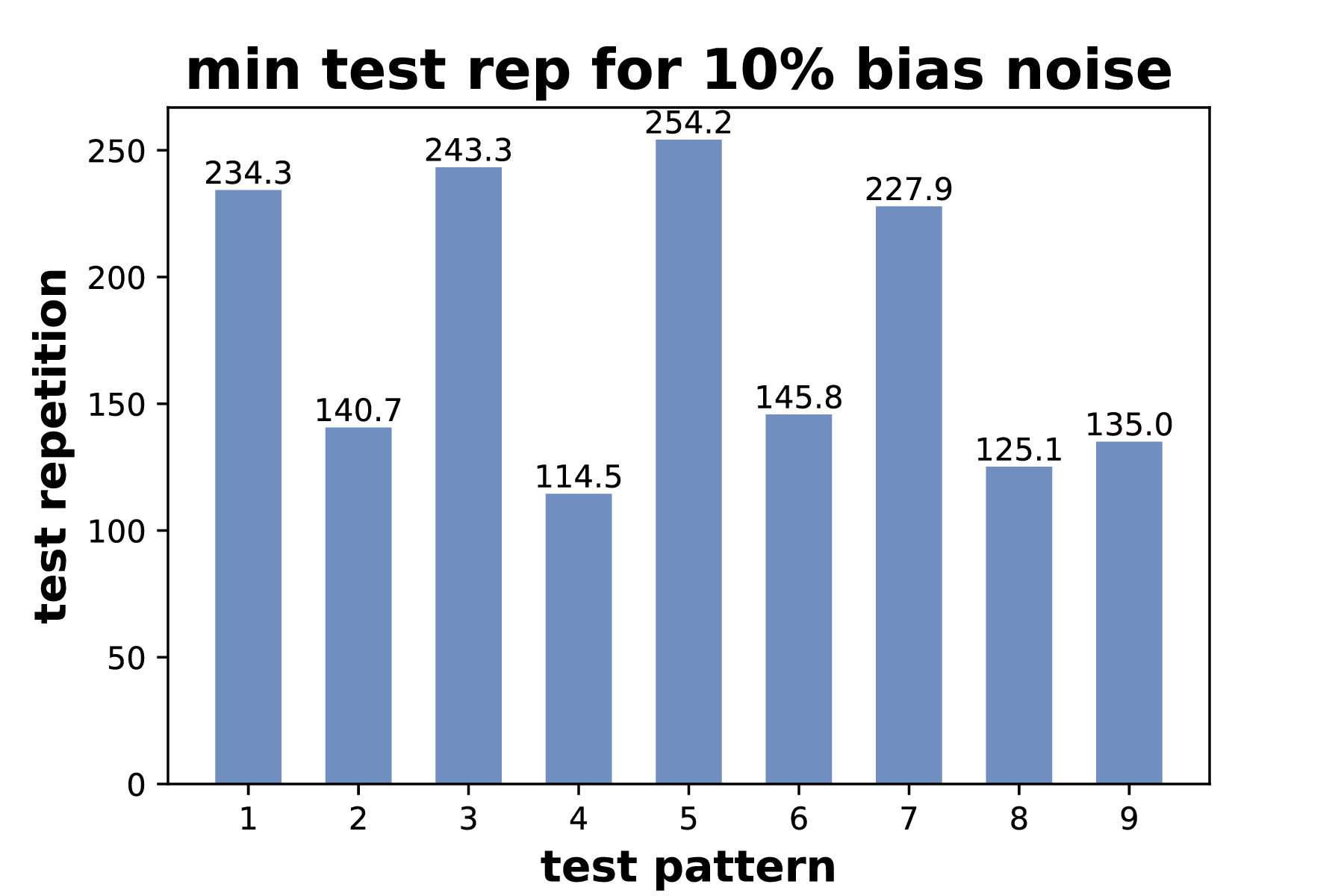}}
    \subfloat[\label{fig:rep20-bi}Test repetition for $20\%$ bias noise]{\includegraphics[width = .45\linewidth]{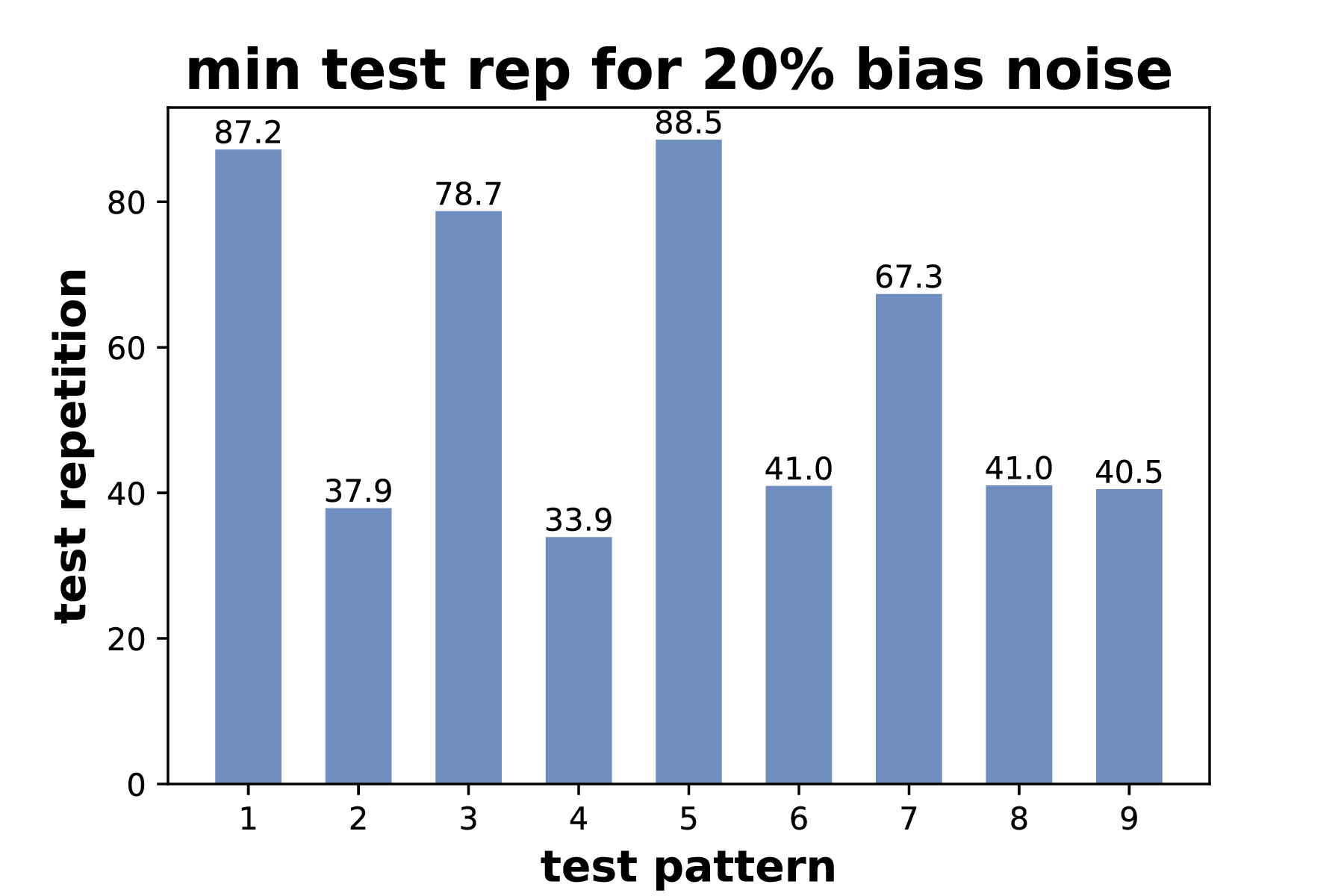}}
    \\
    
    \subfloat[\label{fig:rep10-tr}Test repetition for $10\%$ truncation noise]{\includegraphics[width = .45\linewidth]{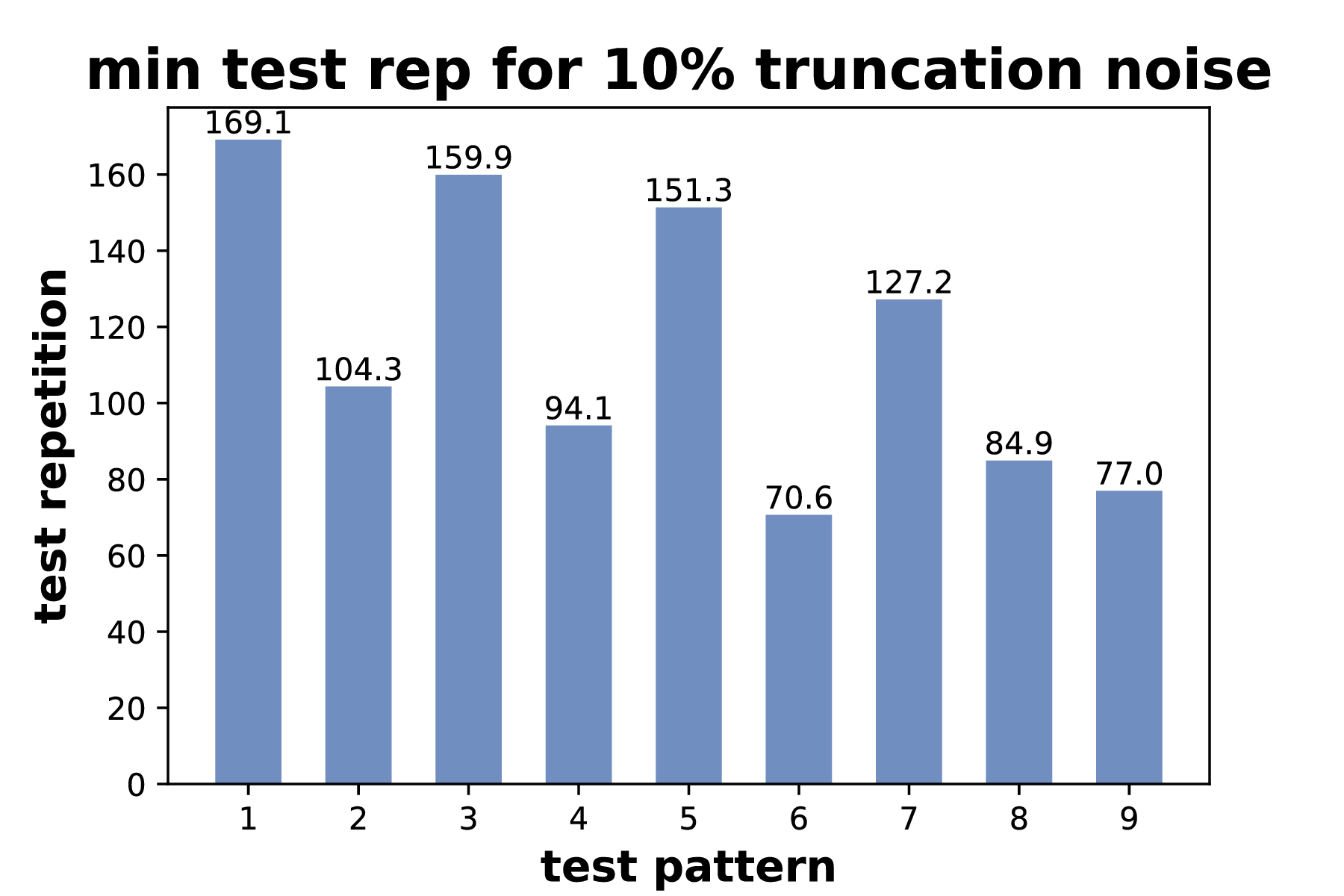}}
    \subfloat[\label{fig:rep20-tr}Test repetition for $20\%$ truncation noise]{\includegraphics[width = .45\linewidth]{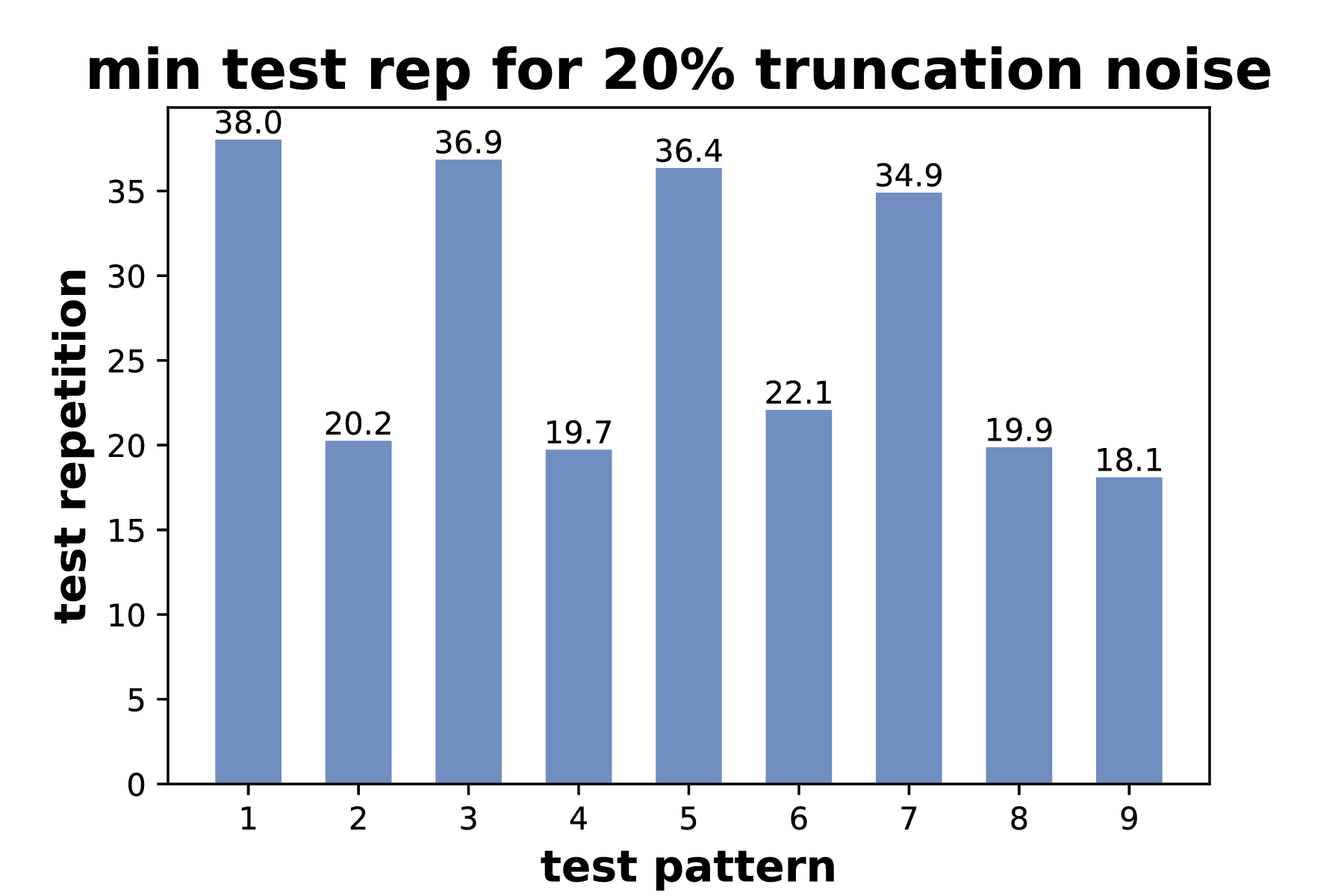}}
    \\
    \caption{Minimal test repetitions for a quantum full adder with errors inserted in one of the CZ gates.}
    \label{fig:pattern2}
\end{figure}

We conduct our experiments on a 4-qubit quantum full adder and a 4-qubit random circuit with the aid of the method in \cite{fault_sim}. The random circuit is shown in Fig. \ref{fig:randcir}. Our experiments mainly include two parts:
\begin{itemize}
    \item 1) We add control errors in all of the CZ gates in each of the two circuits and obtain the minimal test repetition and the corresponding test pattern for different levels of errors. The results of this part are shown in Fig. \ref{fig:pattern1} and Fig. \ref{fig:pattern1-rand}. We find that the minimal test repetition decreases exponentially, which corresponds to the variation of fidelity. Moreover, the test repetitions of random circuit when error is small are prominently larger than those of full adder, which may be explained by the more evenly output distribution of the random circuit. At last, the results show that the best test pattern, i.e., the one with the lowest number of repetition, depends both on the types of the circuit and the strength of the errors. However, for the same circuit, a sufficiently good candidate may exist, e.g., 0111 for the random circuit. 
    
    \item 2) We separately add control noises on different CZ gates in the 4-qubit quantum full adder, one at a time, and find the minimal test repetition for each fault model with certain scales of noises. The results of this part are shown in Fig. \ref{fig:pattern2}. Results shows that the faults in number 1,3,5,7 is harder to detect than other CZ gates. 
\end{itemize}


\section{Conclusion}
In this work we present a fault model for noisy superconducting quantum circuits, including control pulse faults, state leakage faults, decoherence faults and missing gate faults. There are totally $(2m+2)n$ faults that need to be considered in control signals, where $n$ is the number of CZ gates and $m$ is the Fourier terms used in approximation. Using this model, we conducted  fault simulation  on single controlled-Z gate and on certain quantum circuits consisted of single qubit gates and CNOT gates. Simulation results of the noises show that the fidelity decreases exponentially with the noise, and a $10\%$ error in all CZ gates in a 4-qubit full adder takes $34$ test repetitions to detect. Moreover, the noise inserted in one of the CZ gates will lead to either undetectable faults or faults that are detectable within $100$ to $200$ test repetitions. We expect that this work lay a foundation for some future quantum EDA works.

%\bibliographystyle{IEEEtran}
%\bibliography{ref}

\printbibliography

\end{document}